\documentclass[11pt]{article}
\usepackage{amssymb}
\usepackage{graphicx}
\usepackage{amsmath}
\usepackage{makeidx}
\usepackage{indentfirst}

\newcounter{resultnum}[section]
\newcounter{conclusionnum}[section]
\newcounter{conditionnum}[section]
\newcounter{conjecturenum}[section]
\newcounter{examplenum}[section]
\newcounter{exercisenum}[section]
\newcounter{lemmanum}[section]
\newcounter{notationnum}[section]
\newcounter{theoremnum}[section]
\newcounter{definitionnum}[section]
\newcounter{corollarynum}[section]
\newcounter{remarknum}[section]
\newcounter{propositionnum}[section]
\newcounter{acknowledgementnum}[section]
\newcounter{algorithmnum}[section]
\newcounter{axiomnum}[section]
\newcounter{casenum}[section]
\newcounter{claimnum}[section]
\newcounter{summarynum}[section]
\newcounter{problemnum}[section]
\begin{document}

\title{Cyclic and Ekpyrotic Universes in Modified Finsler Osculating Gravity
on Tangent Lorentz Bundles}
\date{December 15, 2012}
\author{\textbf{Panayiotis C. Stavrinos} \thanks{%
pstavrin@math.uoa.gr}  \and  \textsl{\small Department of Mathematics,
University of Athens } \\
\textsl{\small Ilissia, Athens, 15784 Greece} \\
{\qquad} \\
\textbf{Sergiu I. Vacaru} \thanks{%
sergiu.vacaru@uaic.ro,\ http://www.scribd.com/people/view/1455460-sergiu}
\and \textsl{\small Department of the Head of Rector's Office, University "Al. I. Cuza" at Ia\c si, }
\\
\textsl{\small Corpus R, office 323, Alexandru Lapu\c sneanu street, nr. 14,  Ia\c si, Romania, 700054 } }
\maketitle

\begin{abstract}
We consider models of accelerating Universe elaborated for Finsler like
gravity theories constructed on tangent bundles to Lorentz manifolds. In the
osculating approximation, certain locally anisotropic configurations are
similar to those for $f(R)$ gravity. This allows us to generalize a proposal
(by Nojiri, Odintsov and S\'{a}ez--G\'{o}mez, AIP Conf. Proc. \textbf{1458} (2011) 207-221;\ arXiv: 1108.0767) in order to
reconstruct and compare two classes of Einstein--Finsler gravity, EFG, and $%
f(R)$ gravity theories using modern cosmological data and realistic physical
scenarios. We conclude that EFG provides inflation, acceleration and little
rip evolution scenarios with realistic alternatives to standard $\Lambda CDM$
cosmology. The approach is based on a proof that there is a general
decoupling property of gravitational field equations in EFG and modified
theories which allows us to generate off--diagonal cosmological solutions.

\vskip0.2cm


\vskip5pt

\end{abstract}


\section{Introduction and Preliminaries}

Modern cosmology is based on observational data for two accelerating periods
in evolution of Universe, the early--time inflation phase and the late--time
acceleration with dark energy and dark matter effects. Such accelerating
epochs are characterized by a number of differences with large and,
respectively, small values of curvature; quite exotic states of matter at
the "beginning" and "end"; alternative classes of solutions of the
gravitational and matter fields equations, with singularities and possible
anisotropies etc. One of the main exploited ideas is that the Universe is
under cyclic evolution with oscillating equations of matter \cite%
{baum,sahni,dodelson,brevik}. It is considered that it may exist an unified
theory which describes in different limits both the inflation and dark
(energy/matter) periods.

Modified gravity theories were elaborated as unifications and/or
generalizations to Einstein gravity when cyclic cosmology with inflation and
dark energy may be realized (see reviews \cite{nojiri1,nojiri2,capoz}). The
approaches with modified Lagrange density $R\rightarrow f(R,T,...)$, where $%
R $ is a scalar curvature and a functional $f(...)$ is determined by traces
of certain stress--energy tensors, additional torsion fields etc, are
intensively developed in modern literature. Nevertheless, there are other
directions which at first site are alternative to $f(...)$ gravity and
present interest in modern classical and quantum gravity and cosmology. In
this work, there are studied the so--called Finsler gravity and cosmology
(for review of results and critical remarks, see \cite%
{vrflg,vcritic,vcosm1,vcosm2}) and investigate possible connections to $f(R)$
theories. Perhaps, the first model with locally anisotropic inflation in a
Finsler like manner was proposed in \cite{vgontsa}. Several Finsler
cosmology and gravity models where developed in Refs. \cite%
{stavr1,stavr2,mavr1,mavr2,sindoni,visser,vcosm3,vexsol1,veym,vfbr}
following geometric and physical ideas related to quantum gravity and
modified dispersion relations, broken local Lorentz symmetry, nonlinear
symmetries etc.

Our purpose is to prove that a class of Einstein--Finsler gravity (EFG)
theories can be considered as natural candidates for which cyclic cosmology
with inflation and dark energy/matter epoches can be realized.\footnote{%
An EFG model is constructed on a manifold $V$, or its tangent bundle $TV,$
similarly to the GR theory when, roughly speaking, the Levi--Civita
connection $\nabla $ is substituted by a Finsler like metric compatible
connection $\mathbf{D,}$ both defined by the same metric structure $\mathbf{%
g.}$ The second linear connection $\mathbf{D}$ is with nontrivial torsion.}
We shall formulate a reconstruction procedure when alternatively to $f(R)$
gravity \cite{nojiri3} we can extract information on certain locally
anisotropic (Finsler) like gravitational models. It should be noted that the
general relativity (GR) theory and modifications can be re--written in
so--called Finsler like variables which allows us to decouple and integrate
the gravitational and matter field equations in very general forms. This is
possible for nonholonomic/non--integrable $2+2$ splitting of a (pseudo)
Riemannian manifolds enabled with a formal fibred structure. We can mimic a
(pseudo) Finsler geometry on a standard Einstein manifold if we adapt the
constructions with respect to corresponding nonholonomic (non--integrable)
distributions. Such constructions are rather formal but allow us to
formulate a general geometric method of constructing exact solutions both in
GR and $f(R,T,...)$ modifications, when the generic off-diagonal metrics and
various types of connection and frame variables depend on all coordinates
via generating and integration functions and parameters.

Let us summarize some key ideas on modifications/ generalizations of the
Einstein gravity on (co) tangent bundles. In such theories, the fundamental
geometric objects (metrics, frames and connections) depend on velocity/
momentum type variables which can be interpreted as fiber like coordinates.
The geometric constructions are derived from a nonlinear quadratic element $%
ds^{2}=F^{2}(x,y),$ where $x=(x^{i})$ and $y=(y^{a})$ are coordinates on a
tangent a bundle $T\mathbf{V}$ to a Lorentz manifold $\mathbf{V}$. Such base
manifolds are necessary if we wont to get in a limit the standard GR theory.%
\footnote{%
The well known pseudo--Riemannian geometry consists a particular case with
quadratic element $ds^{2}=g_{ij}(x)dx^{i}dx^{j},$ when $y^{i}\sim dx^{i}.$}
The value $F$ is called the fundamental/generating Finsler function.
Usually, certain homogeneity conditions on $y$ are imposed and the
nondegenerated Hessian
\begin{equation}
\tilde{g}_{ab}:=\frac{1}{2}\frac{\partial ^{2}F}{\partial y^{a}\partial y^{b}%
}  \label{hess}
\end{equation}%
is considered as the fiber metric. Gravity theories with anisotropies on $T%
\mathbf{V}$ can not be determined only by $F,$ or $\tilde{g}_{ab}.$ This is
very different from GR and a class of $f(R,T,...)$ theories (on
pseudo--Riemannian spaces) when the geometric and physical models are
completely defined by data the metric structure. To construct a Finsler
geometry/gravity theory we need a triple $(F:\mathbf{g,N,D})$ of fundamental
geometric objects:\ the total metric structure, $\mathbf{g},$ the nonlinear
connection (N--connection) structure, $\mathbf{N}, $ and the distinguished
connection structure, $\mathbf{D},$ which is adapted to $\mathbf{N}$. There
are necessary additional assumptions how such objects are defined by $F,$
which types of compatible or noncompatible linear connections are involved,
how the corresponding curvature, torsion and nonmetricity fields must be
computed. For realistic physical models, the experimental/observational
effects can be analyzed via a so--called osculating approximation (see
examples in \cite{stavr1,stavr2})
\begin{equation}
\tilde{g}_{ij}=\tilde{g}_{ij}(x,y(x))=\frac{1}{2}\frac{\partial ^{2}F}{%
\partial y^{a}\partial y^{b}}(x,y(x)).  \label{osc}
\end{equation}%
For a review of Finsler geometry for physicists, see Refs. \cite%
{vrflg,vcosm2} and references therein.

The paper is structured as follows. In section \ref{sefg}, we summarize the
necessary results on Einstein--Finsler and $f(R)$ modified gravity and show
how such theories can modelled by nonholonomic distributions and/or
off--diagonal metrics in GR. Section \ref{sccu} is devoted to models of
conformal cyclic universes in EFG. We consider FLRW metrics subjected to
nonholonomic constraints and analyze consequences of locally anisotropic
models. A procedure of reconstructing EFG theories is provided. There are
studied several models with effective anisotropic fluids. Scenarios of
ekpyrotic and little rip cosmology governed by fundamental Finsler functions
in osculating approximation are studied in section \ref{stbc}. The
anholonomic deformation method is applied for generating cosmological
solutions in EFG. Conclusions are drawn in section \ref{sconcl}.

\section{ Canonical EFG \& $f(R)$ Modifications}

\label{sefg}We consider a four dimensional (4--d) Lorentz manifold $\mathbf{V%
}$ in GR modelled as a pseudo--Riemannian space enabled with a metric $\ ^{h}%
\mathbf{g}=\{g_{ij}(x^{i})\}$ of local signature $(+,+,+,-).$ Physically
motivated Finsler generalizations to metrics and other geometric objects
depending anisotropically of velocity/momentum type variables $y^a$ can be
constructed on tangent bundle $T\mathbf{V.}$ A (pseudo) Finsler geometry is
characterized by its fundamental (equivalently, generating) functions when
certain homogeneity conditions are imposed, $F(x^{i},\beta y^{j})=\beta
F(x^{i},y^{j}),$ for any $\beta >0,$ and $\det |\tilde{g}_{ab}|\neq 0,$ see
Hessian (\ref{hess}) which defines a so--called vertical metric. In standard
approaches, the matrix $\tilde{g}_{ab}$ is considered positively definite,
but this condition has to be dropped in (pseudo) Finsler geometry
(hereafter, we shall omit the term "pseudo").

\subsection{The "triple" of fundamental geometric objects}

A canonical model of Finsler--Cartan geometry is completely determined by $%
F, $ and $\tilde{g}_{ij}$, up to necessary classes of frame/coordinate
transform $e^{\alpha ^{\prime }}=e_{\ \alpha }^{\alpha ^{\prime
}}(x,y)e^{\alpha },$ following such assumptions on the triple $\left( F:%
\mathbf{N,g,D}\right) $ of fundamental geometric objects (see details in
\cite{vrflg,vcosm2}):

There is a canonical nonlinear connection (N--connection) structure,
\begin{equation}
\mathbf{N}:\ T\mathbf{TV}=h\mathbf{TV}\oplus v\mathbf{TV,}  \label{whit}
\end{equation}%
which can be introduced as a nonholonomic (non--integrable/ anholonomic)
distribution with horizontal (h) and vertical (v) splitting.\footnote{\label%
{fnotenc}To define the canonical N--connection we follow a
geometric/variational principle for an effective regular Lagrangian $L=F^{2}$
and action $S(\tau )=\int\limits_{0}^{1}L(x(\tau ),y(\tau ))d\tau ,%
\mbox{
for }y^{k}(\tau )=dx^{k}(\tau )/d\tau $. The Euler--Lagrange equations $%
\frac{d}{d\tau }\frac{\partial L}{\partial y^{i}}-\frac{\partial L}{\partial
x^{i}}=0$ are equivalent to the \textquotedblright nonlinear
geodesic\textquotedblright\ (equivalently, semi--spray) equations $\frac{%
d^{2}x^{k}}{d\tau ^{2}}+2\tilde{G}^{k}(x,y)=0.$ The canonical coefficients $%
\mathbf{\tilde{N}}=\{\tilde{N}_{i}^{a}\}$ are computed $\tilde{N}_{j}^{a}:=%
\frac{\partial \tilde{G}^{a}(x,y)}{\partial y^{j}},\ \tilde{G}^{k}=\frac{1}{4%
}\tilde{g}^{kj}\left( y^{i}\frac{\partial ^{2}L}{\partial y^{j}\partial x^{i}%
}-\frac{\partial L}{\partial x^{j}}\right) $.
\par
In our works, we put "tilde" on symbols (for instance, $\mathbf{\tilde{e}}%
_{\alpha }$) if the constructions are performed with respect to bases
determined by $\tilde{N}_{j}^{a}$. "Tilde" will be omitted for general
N--adapted $h$- $v$--splitting when variables mix each to others.} We get an
integrable/holonomic frame configuration if $W_{\alpha \beta }^{\gamma }=0$.
Under frame transforms, the coefficient formulas transform into equivalent
ones for arbitrary sets of coefficients $\mathbf{N}=\{\mathbf{N}_{i^{\prime
}}^{a^{\prime }}=e_{\ a}^{a^{\prime }}e_{i^{\prime }}^{\ i}\tilde{N}%
_{i}^{a}\}.$ We can adapt the geometric constructions via "N--elongated"
(co)frame structures,
\begin{eqnarray}
\mathbf{e}_{\nu } &=&(\mathbf{e}_{i},e_{a}),\ \mathbf{e}_{i}=\frac{\partial
}{\partial x^{i}}-N_{i}^{a}(u)\frac{\partial }{\partial y^{a}}\mbox{ and
}e_{a}=\frac{\partial }{\partial y^{a}};  \label{dder} \\
\mathbf{e}^{\mu } &=&(e^{i},\mathbf{e}^{a}),\ e^{i}=dx^{i}\mbox{ and }%
\mathbf{e}^{a}=dy^{a}+N_{i}^{a}(u)dx^{i}.  \label{ddif}
\end{eqnarray}%
In general, such frames are nonholonomic because
\begin{equation}
\lbrack \mathbf{e}_{\alpha },\mathbf{e}_{\beta }]=\mathbf{e}_{\alpha }%
\mathbf{e}_{\beta }-\mathbf{e}_{\beta }\mathbf{e}_{\alpha }=W_{\alpha \beta
}^{\gamma }\mathbf{w}_{\gamma }  \label{anhrel}
\end{equation}%
with $W_{ia}^{b}=\partial _{a}N_{i}^{b}$ and $W_{ji}^{a}=\Omega _{ij}^{a}=%
\frac{\partial N_{i}^{a}}{\partial x^{j}}-\frac{\partial N_{j}^{a}}{\partial
x^{i}}+N_{i}^{b}\frac{\partial N_{j}^{a}}{\partial y^{b}}-N_{j}^{b}\frac{%
\partial N_{i}^{a}}{\partial y^{b}}.$

A canonical metric structure on $\mathbf{TV,}$ can be introduced using data $%
\left( {g}_{ij},\mathbf{e}_{\alpha }\right) ,$ with $\mathbf{e}_{\alpha
^{\prime }}=e_{\ \alpha ^{\prime }}^{\alpha }\mathbf{e}_{\alpha }$ and $%
\mathbf{g}_{\alpha ^{\prime }\beta ^{\prime }}=e_{\ \alpha ^{\prime
}}^{\alpha }e_{\ \beta ^{\prime }}^{\beta }\mathbf{\tilde{g}}_{\alpha \beta
},$
\begin{equation}
\mathbf{g}=h\mathbf{g}\oplus v\mathbf{g}=\ g_{ij}(x,y)\ e^{i}\otimes e^{j}+\
h_{ab}(x,y)\ \mathbf{e}^{a}\otimes \mathbf{e}^{b}.  \label{dm}
\end{equation}

The third fundamental geometric object in a Finsler geometry is the
distinguished connections (d--connection) $\mathbf{D=\{\Gamma }_{\beta
\gamma }^{\alpha }\mathbf{\}=}(hD,vD)$ which by definition is adapted to the
N--connection structure, i.e. preserves the nonholonomic h-v--splitting (\ref%
{whit}). We can model physically viable "almost" standard models (see
discussions and critical remarks in \cite{vcritic,vrflg}) for d--connections
which are compatible with the metric structure, $\mathbf{Dg=0}$ and
completely defined by data $(\mathbf{g},\mathbf{N}).$ Let us consider how
such a canonical d--connection\ $\mathbf{D}$ can be constructed.\footnote{%
Using N--adapted differential forms and the d--connection 1--form is $%
\mathbf{\Gamma }_{\ \beta }^{\alpha }=\mathbf{\Gamma }_{\ \beta \gamma
}^{\alpha }\mathbf{e}^{\gamma },$ we compute the torsion, $\mathcal{T}%
^{\alpha },$ and curvature 2--forms, $\mathcal{R}_{~\beta }^{\alpha },$ $%
\mathcal{T}^{\alpha }:=\mathbf{De}^{\alpha }=d\mathbf{e}^{\alpha }+\mathbf{%
\Gamma }_{\ \beta }^{\alpha }\wedge \mathbf{e}^{\beta },\ \mathcal{R}%
_{~\beta }^{\alpha }:=\mathbf{D\Gamma }_{\ \beta }^{\alpha }=d\mathbf{\Gamma
}_{\ \beta }^{\alpha }-\mathbf{\Gamma }_{\ \beta }^{\gamma }\wedge \mathbf{%
\Gamma }_{\ \gamma }^{\alpha }=\mathbf{R}_{\ \beta \gamma \delta }^{\alpha }%
\mathbf{e}^{\gamma }\wedge \mathbf{e}^{\delta }.$ The torsion coefficients $%
\mathbf{T}_{\ \beta \gamma }^{\alpha }=\{T_{\ jk}^{i},T_{\ ja}^{i},T_{\
ji}^{a},T_{\ bi}^{a},T_{\ bc}^{a}\}$ are $T_{\ jk}^{i}=L_{\ jk}^{i}-L_{\
kj}^{i},\ T_{\ ja}^{i}=-T_{\ aj}^{i}=C_{\ ja}^{i},\ T_{\ ji}^{a}=\Omega _{\
ji}^{a},\ T_{\ bi}^{a}=\frac{\partial N_{i}^{a}}{\partial y^{b}}-L_{\
bi}^{a},\ T_{\ bc}^{a}=C_{\ bc}^{a}-C_{\ cb}^{a};$\ for N--adapted
coefficients $\mathbf{R}_{\ \beta \gamma \delta }^{\alpha },$ see \cite%
{vrflg,vcosm2}.
\par
In our former works the symbol $\widehat{\mathbf{D}}$ was used for the
canonical d--connection. We shall omit "hats" if that will not result in
ambiguities. The N--adapted coefficients of $\mathbf{D}$ can be computed in
the form $\mathbf{\Gamma }_{\ \alpha \beta }^{\gamma }=\left(
L_{jk}^{i},L_{bk}^{a},C_{jc}^{i},C_{bc}^{a}\right) ,$
\begin{eqnarray*}
L_{jk}^{i} &=&\frac{1}{2}g^{ir}\left( \mathbf{e}_{k}g_{jr}+\mathbf{e}%
_{j}g_{kr}-\mathbf{e}_{r}g_{jk}\right) ,\ \widehat{C}_{bc}^{a}=\frac{1}{2}%
h^{ad}\left( e_{c}h_{bd}+e_{c}h_{cd}-e_{d}h_{bc}\right) , \\
L_{bk}^{a} &=&e_{b}(N_{k}^{a})+\frac{1}{2}h^{ac}\left( e_{k}h_{bc}-h_{dc}\
e_{b}N_{k}^{d}-h_{db}\ e_{c}N_{k}^{d}\right) ,\ C_{jc}^{i}=\frac{1}{2}%
g^{ik}e_{c}g_{jk}.
\end{eqnarray*}%
} By definition, such a connection is with zero $h$- and $v$-torsions, $T_{\
jk}^{i}=0$ and $T_{\ bc}^{a}=0.$ There are also nontrivial torsion
coefficients. In Finsler geometry, it is also possible to introduce in
standard form the Levi--Civita connection $\nabla =\{\Gamma _{\beta \gamma
}^{\alpha }\}$ (as a unique one which is metric compatible and zero torsion)
but it does not preserve under parallelism the N--connection splitting (\ref%
{whit}). There is a canonical distortion relation
\begin{equation}
\mathbf{D=}\nabla +\mathbf{Z},  \label{dcdc}
\end{equation}%
where $\mathbf{D,}$ $\nabla $ and the distortion tensor $\mathbf{Z}$ are
uniquely defined by the same metric structure $\mathbf{g.}$

Finally, we note that there is an important argument to work with $\mathbf{D}
$ with respect to N--adapted frames (\ref{dder}) and (\ref{ddif}) even in
GR. This is possible for a conventional nonholonomic 2+2 spitting. The
priority is that $\mathbf{D}$ allows to decouple and integrate the
gravitational field equations in very general forms \cite{vexsol1,veym,vfbr}
(see section \ref{stbc}). Such solutions define standard Lorentz/Einstein
manifolds if we constrain at the end the generating/integration functions in
such a form that $\mathbf{Z}=0$ which mean zero torsion. For such
configurations, we get $\mathbf{D}_{\mid \mathbf{T=0}}\mathbf{=}\nabla $ in
N--adapted form, see (\ref{dcdc}).

\subsection{The EFG field equations and $\ f(R)$ gravity}

A Einstein--Finsler gravity theory (EFG) is constructed for a Finsler like
d--connection $\mathbf{D}$ following standard geometric and/or variation
rules as in GR. In this work, we shall use only the canonical d--connection.
Such models can be constructed, for instance, on a Lorentz manifold $\mathbf{%
V}$ (considering an 2+2 splitting; in this case, we introduce Finsler like
variables in GR, when $\mathbf{T=0}$ is considered as nonholonomic
constraint for $\mathbf{D\rightarrow \nabla ,}$ or on its tangent bundle $T%
\mathbf{V.}$ In the second case, we can relate the constructions to certain
models of Finsler--Cartan gravity (via nonholonomic deformations, we can
transform $\mathbf{D}$ into the Cartan connection for Finsler space)%
\footnote{%
We use such conventions on indices $\alpha =(i,a):$ For a 2+2 splitting on $%
\mathbf{V,}$ in GR, we consider $i,j,...=1,2$ and $a,b,...=3,4.$ On $T%
\mathbf{V,}$ $i,j,...=1,2,3,4$ and $a,b,...=5,6,7,8.$ Here we note that
formal integrations of gravitational field equations in general forms are
possible for splitting of type $2+2+2+2+...$ and/or $3+2+...$}.

The Ricci tensor, $Ric=\{\mathbf{R}_{\ \beta \gamma }:=\mathbf{R}_{\ \beta
\gamma \alpha }^{\alpha }\},$ of $\mathbf{D=}(hD,vD),$ splits into certain $%
h $- and $v$-components $(R_{ij},R_{ai},R_{ia},R_{ab}).$The scalar curvature
is
\begin{equation}
\ ^{F}R:=\mathbf{g}^{\beta \gamma }\ \mathbf{\mathbf{R}}_{\beta \gamma
}=g^{ij}R_{ij}+h^{ab}R_{ab}=\tilde{R}+\breve{R}.  \label{riccifs}
\end{equation}%
The gravitational field equations can be postulated in geometric form
(and/or derived via N--adapted variational calculus),
\begin{equation}
\mathbf{R}_{\ \beta \delta }-\frac{1}{2}\mathbf{g}_{\beta \delta }\ \ ^{F}R=%
\mathbf{\Upsilon }_{\beta \delta }.  \label{cdeins}
\end{equation}%
Such equations transform into the Einstein's ones in GR if $\mathbf{\Upsilon
}_{\beta \delta }\rightarrow T_{\beta \delta }$ and $\mathbf{D}\rightarrow
\nabla .$ So, we can consider necessary N--adapted variations of actions for
scalar, electromagnetic, spinor etc fields in order to derive certain the
matter source in (\ref{cdeins}). The solutions of the field equations in EFG
are with nontrivial torsion induced by $(\mathbf{g},\mathbf{N}).$ The
Levi--Civita (zero torsion) conditions with respect to N--adapted frames are
\begin{equation}
L_{aj}^{c}=e_{a}(N_{j}^{c}),C_{jb}^{i}=0,\ \Omega _{\ ji}^{a}=0.
\label{lccond}
\end{equation}%
If such constraints are imposed at the end after we have constructed certain
classes of solutions of (\ref{cdeins}) on a 4--d manifold and for physically
motivated sources, we generate (in general, off--diagonal) solutions in GR.

It is not clear what types of sources \ $\mathbf{\Upsilon }_{\beta \delta }$
should be considered in EFG models on $T\mathbf{V.}$ We note that, in
general, such tensors are not symmetric because $\mathbf{R}_{\ \beta \delta
} $ is not symmetric. It reflects the nonholonomic and locally anisotropic
character of Finsler gravity theories. For certain toy models, we can
approximate \ $\mathbf{\Upsilon }_{\beta \delta }$ to a cosmological
constant with possible locally anisotropic polarizations depending on $%
\left( x^{i},y^{a}\right) .$ The fundamental Finsler function $F(x,y)$ is
encoded into geometric objects of (\ref{cdeins}) and solutions of such PDE.
\ In general, we can not "see" explicit dependencies on $F(x,y)$ because of
principles of generalized covariance and relativity: arbitrary
frame/coordinate transforms on $\mathbf{V}$ and $T\mathbf{V}$ mix the
variables. Fixing a system of local coordinates, we can measure
experimentally on $h$--subspace only an osculating (pseudo) Riemannian
metric (\ref{osc}). Mathematically, we can construct exact solutions of (\ref%
{cdeins}) for 8-d metrics (\ref{dm}) but to verify possible physical
implications in direct form is possible only for the $h$--components.

EFG theories on $\mathbf{V,}$ or $T\mathbf{V,}$ are explicit examples of
modified GR. In a different manner, such theories of gravity are constructed
for modified Lagrange densities when $R\rightarrow f(R,T,...),$ see reviews
of results in are \cite{nojiri1,nojiri2,capoz}. For instance, the
corresponding field equations with $f(R)$ can be written as effective
Einstein fields equations
\begin{equation}
R_{\mu \nu }-\frac{1}{2}Rg_{\mu \nu }=8\pi \ ^{ef}G\ ^{ef}T_{\mu \nu },
\label{meinst}
\end{equation}%
for $\ ^{ef}T_{\mu \nu }=(\partial _{R}f)^{-1}[\frac{1}{2}[f-R\partial
_{R}f]g_{\mu \nu }-(g_{\mu \nu }\nabla ^{\alpha }\nabla _{\alpha }-\nabla
_{\mu }\nabla _{\nu })\partial _{R}f]$ and $\ ^{ef}G=(\partial _{R}f)^{-1}$.

Using observational/ experimental data, we can measure some geometric
configurations determined by functional dependencies of type $%
g_{ij}[F,f,...].$ In general, it may be not clear if such a metric is a
solution of (\ref{cdeins}) or (\ref{meinst}), i.e. we can not say exactly
what kind of modifications of GR would result in dark energy/matter effects.
If a fundamental Finsler function $F$ is considered as a nonholonomic
distribution on a pseudo--Riemannian space $V,$ or $TV,$ modifications of
type $R\rightarrow f(R)$ result in $(F;\ ^{F}R)$ $\rightarrow f(R,F)\simeq f(%
\tilde{R})\simeq f(\ ^{F}R),$ see formulas (\ref{cdeins}) for the Finsler
curvature scalar and its $h$- and $v$--splitting. In general, such
modifications can be performed for any model of Finsler spacetime geometry
with scalar curvature $\ ^{F}R$. A physically realistic theory closed to GR
and MG can be constructed in the simplest way for the scalar curvature $%
\tilde{R}$ of the Cartan d--connection. For a prescribed N--connection
structure, we can define via nonholonomic deformations (\ref{dcdc})
functional dependencies of type $\ \tilde{R}(R)$ and, inversely, $R(\tilde{R}%
).$ In general, EFG and $f(...)$ are different modifications of GR described
by different Lagrange densities and derived field equations. For certain
conditions, we can transform a class of solutions of (\ref{cdeins}) into (%
\ref{meinst}), and inversely, using frame/coordinate transforms. It is known
a reconstruction technique \cite{nojiri3} allows us to recover data for $f(R)
$ using observational data from modern cosmology. In this work, we formulate
a similar approach for extracting data for EFG theories and speculate on
conditions when we can distinguish a Finsler configuration from a $f$%
--modification.

\section{Conformal Cyclic Finsler Like Universes}

\label{sccu}In this section, we show how to construct cycling universes in
models with $f(R,F)\simeq f(\tilde{R})$ when the theory is determined by
actions of type%
\begin{equation}
S=\int d^{4}x\sqrt{|g_{ij}|}\left[ \frac{f(\tilde{R})}{2\kappa ^{2}}+\ ^{m}L%
\right] \mbox{\ and/or\  }S=\int d^{4}x\sqrt{|g_{ij}|}\left[ \frac{\tilde{R}%
(R)}{2\kappa ^{2}}+\ ^{m}L\right] ,  \label{modact}
\end{equation}%
where $\kappa $ is the gravitational constant, $\ ^{m}L$ is the Lagrangian
for density and the osculating approximation (\ref{osc}) is encoded in $%
g_{ij},\tilde{R}$ and (for simplicity, we shall consider models with
functional dependence on metric for matter Lagrangians) $\ ^{m}L.$

A scalar field $\chi $ can be introduced via conformal transforms $\mathbf{g}%
_{\alpha \beta }\rightarrow e^{-\chi (x)}\mathbf{g}_{\alpha \beta }$ with
respect to N--elongated frames (\ref{dder}) and (\ref{ddif}).\footnote{%
We do not write $\phi $ for scalar fields as, for instance, in \cite{nojiri3}
but use $\chi $ because in our works $\phi $ is considered as a generating
function for constructing off--diagonal solutions \cite{vexsol1,veym,vfbr}
(see also next section).} Denoting
\begin{equation*}
_{\chi }\tilde{R}=e^{\chi }\{\tilde{R}+3\mathbf{g}^{ij}[(\mathbf{D}_{i}\chi
)(\mathbf{D}_{j}\chi )-\frac{1}{2}(\mathbf{e}_{i}\chi )(\mathbf{e}_{j}\chi
)]\},
\end{equation*}
we get a new action
\begin{equation}
S=\int d^{4}xe^{-2\chi }\sqrt{|g_{ij}|}\left[ \frac{1}{2\kappa ^{2}}f(_{\chi
}\tilde{R},\chi )+\ ^{m}L\right] .  \label{act}
\end{equation}%
Choosing some "proper" functionals $\mathbf{P}(\chi )$ and $\mathbf{Q}(\chi
) $ when the equation
\begin{equation}
\partial _{\chi }\mathbf{P}(\chi )\ \tilde{R}+\partial _{\chi }\mathbf{Q}%
(\chi )=0  \label{geneq}
\end{equation}
can be solved as $\chi =\chi (\tilde{R}),$ we can express $\ f(\tilde{R})=%
\mathbf{P[}\chi (\tilde{R})]\ \tilde{R}+\mathbf{Q[}\chi (\tilde{R})].$

For "pure" Finsler modifications, the functionals (\ref{geneq}) are of type $%
\ \tilde{R}=\tilde{R}(R)$ with determined in nonholonomic form via
distortions (\ref{dcdc}), where $R$ is the scalar curvature of $\nabla .$ In
such cases, actions of type (\ref{act}) are parameterized in the form
\begin{equation}
S=\int d^{4}xe^{-2\chi }\sqrt{|g_{ij}|}\left[ \frac{1}{2\kappa ^{2}}\ \tilde{%
R}(R,\chi )+\ ^{m}L\right] ,  \label{actfinsler}
\end{equation}%
when
\begin{equation}
\ \tilde{R}(R)=\mathbf{\tilde{P}[}\chi (R)]R+\mathbf{\tilde{Q}[}\chi (R)].
\label{functa}
\end{equation}

We conclude that an EFG theory with osculating approximation is a variant of
modified gravity with $f\rightarrow \tilde{R}$. Such theories can be studied
by the similar methods if we work with respect to N--adapted frames. In
former approaches to $f(R,T)$ modifications of gravity, such "preferred"
bases and geometric constructions were not considered.

\subsection{Reconstructing EFG theories}

There are three possibilities to introduce locally anisotropic scalar fields
in a $\ \tilde{R}$--modified theory: 1) via conformal transforms as in (\ref%
{act}); 2) with certain compactification of extra dimensions of models on $T%
\mathbf{V;}$ 3) postulating scalar field Lagrangians with operators $\mathbf{%
e}_{i}$ and the canonical d--connection $\mathbf{D}_{i}.$ Models of type
1)--3) mutually transform from one into another one under frame/coordinate
transforms.

Let us introduce a system of local coordinates $(x^{1},x^{2},x^{3}=t,x^{4}),$
with time like coordinate $t,$ on $\mathbf{V,}$ and osculating approximation
of Finsler type $\tilde{g}_{ij}=\tilde{g}_{ij}(t,y^{a}(t))$ and $\tilde{N}%
_{i}^{a}=\tilde{N}_{i}^{a}(t,y^{b}(t))$ with "non--tildes" with respect to
arbitrary frames. On $h$--subspace, we can consider a FLRW ansatz of type (%
\ref{dm}) when
\begin{equation}
ds^{2}=-(e^{3})^{2}+\tilde{a}^{2}(t+\widehat{t})\left[ (\mathbf{e}^{1})^{2}+(%
\mathbf{e}^{2})^{2}+(e^{4})^{2}\right] +\{\mbox{unknown v--components}\},
\label{periodf}
\end{equation}%
with
\begin{equation}
\tilde{a}(t)=\exp [H_{0}t+b(t)],  \label{cyclfact}
\end{equation}%
where the constant $H_{0}>0$ and $\widehat{t}$ is the period. We use tilde
on $a$ in order to emphasize that this value contains certain Finsler
information. Such a generic off--diagonal metric is considered to result in
physical identical scenarios at $t$ and $t+\widehat{t}.$ For instance, we
can take $b(t)=\ ^{0}b\cos (2\pi t/\widehat{t})$ when the condition $\
^{0}b<H_{0}$ results in monotonic expansion. We emphasize that the $h$--part
of quadratic element on $T\mathbf{V}$ is the same as in the 4--d flat
cosmology. Nevertheless, such models are with osculating dependence on $%
t,y^{a}(t)$ and contain certain information on a possible Finsler spacetime
structure with N--elongated partial derivatives of type (\ref{dder}). To
show this we consider a locally anisotropic scalar--tensor model. The action
is taken following the possibility 3) above,
\begin{equation}
S=\int d^{4}x\sqrt{|g_{ij}|}\left[ \frac{\tilde{R}}{2\kappa ^{2}}-\frac{1}{2}%
\omega (\chi )g^{ij}(\mathbf{e}_{i}\chi )(\mathbf{e}_{j}\chi )-\mathcal{V}%
(\chi )+\ ^{m}L\right] ,  \label{actf1}
\end{equation}%
for some functionals $\omega (\chi )$ and $\mathcal{V}(\chi )$ on $\chi .$
If both values are determined by a single function $\varsigma (\chi ),$ we
express%
\begin{equation*}
\omega (\chi )=-(2/\kappa ^{2})\partial _{\chi \chi }^{2}\varsigma (\chi )%
\mbox{ \ and \ }\mathcal{V}(\chi )=\kappa ^{-2}\{3[\partial _{t}\varsigma
(\chi )]^{2}+\partial _{\chi \chi }^{2}\varsigma (\chi )\}
\end{equation*}%
for a cosmological solution with $h$--part of FLRW type, when $\chi =t$ and
the Hubble function $H=\partial _{t}\varsigma .$\footnote{%
We note that for cosmological models, the rule of N--elongation of partial
derivatives (\ref{dder}) for $A(x(t),y(x(t)))$ is
\begin{equation*}
\partial _{t}A\rightarrow \mathbf{e}_{t}A=\partial _{t}A-N_{t}^{a}\frac{%
\partial A}{\partial y^{a}},
\end{equation*}%
which points to possible contributions from the fibred structure.
Considering a scaling factor $a(t)$ in a nontrivial nonholonomic (Finsler)
background, we introduce $H(t):=\left[ \mathbf{e}_{t}a(t)\right] /a(t).$}

For instance, if we parameterize the cycling factor (\ref{cyclfact}) $\
\tilde{a}(t)=\exp [\varsigma (\chi )]=\exp [\varsigma (t)]$, we recover a
scalar--tensor model with spherical symmetry on $hT\mathbf{V}$ and reproduce
the cyclic universe via formulas which are similar to that presented in Sec.
II of \cite{nojiri3},
\begin{eqnarray*}
\omega (\chi ) &=&(2\ ^{0}b/\kappa ^{2})(2\pi \widehat{t}^{-1})^{2}\cos
(2\pi \chi \widehat{t}^{-1}), \\
\mathcal{V}(\chi ) &=&\kappa ^{-2}\{3[H_{0}-(2\pi \ ^{0}b\widehat{t}%
^{-1})\sin (2\pi \chi \widehat{t}^{-1})]^{2}-(2\pi \widehat{t}^{-1})^{2}\cos
(2\pi \chi \widehat{t}^{-1})\}.
\end{eqnarray*}%
It is not possible to say to much on underling Finsler structure using such
solutions. We can only conclude that a model with h--scalar $\tilde{R}$ may
be with cyclic behaviour. In order to extract more information, we should
perform a rigorous analysis how $F$ modifies the GR.

We apply the methods of reconstructing modified gravity theories \cite%
{nojiri1,nojiri2,nojiri4,nojiri3} reformulated for models of type (\ref%
{modact}) with $\tilde{R}(R)$ and functionals $\mathbf{\tilde{P}[}\chi ]R+%
\mathbf{\tilde{Q}[}\chi ]$ (\ref{functa}) and scaling factor parameterized
in the form $\tilde{a}(t)=\ ^{0}a\ e^{\varpi (t)},\ ^{0}a=const.$ The
solutions for $\mathbf{\tilde{P}}$ and $\mathbf{\tilde{Q}}$ can be found
expanding such values in Fourier series and defining certain recurrent
formulas for coefficients expressed in terms of $H_{0}$ and $\ ^{0}b.$ For
simplicity, we provide the equation for $\mathbf{\tilde{P}}$ when the matter
can be neglected,%
\begin{equation*}
\partial _{\chi }\left[ \varpi (\chi )/\sqrt{|\mathbf{\tilde{P}}(\chi )|}%
\right] =-\frac{1}{2}[\partial _{\chi \chi }\mathbf{\tilde{P}}(\chi )]/(%
\sqrt{|\mathbf{\tilde{P}}(\chi )|})^{3}.
\end{equation*}%
The equation for $\varpi (\chi )$ is
\begin{equation}
\partial _{\chi }\varpi =-\frac{1}{2}\sqrt{|\mathbf{\tilde{P}}|}\int d\chi
(\partial _{\chi \chi }\mathbf{\tilde{P})/}(\sqrt{|\mathbf{\tilde{P}}|})^{3}.
\label{gensol2}
\end{equation}%
We can consider a functional $\mathbf{\tilde{P}}$ in (\ref{geneq}) for
certain values of $\tilde{R}$ as a generating function for solutions of (\ref%
{gensol2}). Here we provide a well known example with divergences
corresponding to $a(t_{0})=0$ which can be identified with a moment of
singularity for a big bang, or crunch effect. \ Fixing $\mathbf{\tilde{P}}%
=P_{0}\left[ \cos (P_{1}\chi )\right] ^{4},$ for some constants $P_{0}$ and $%
P_{1},$ we express the last equation as
\begin{equation*}
\partial _{\chi }\varpi =\varpi _{0}\left[ \cos (P_{1}\chi )\right]
^{2}+2P_{1}\left[ \sin (2P_{1}\chi )-\tan (2P_{1}\chi )\right] ,
\end{equation*}%
for an integration constant $\varpi _{0}.$ The term with $\tan (2P_{1}\chi )$
results in divergences of as it is well known for cyclic scenarios and
ekpyrotic effects. This allows us to reproduce partially various models of
modified gravity including EFG. If the Levi--Civita condition $\mathbf{D}%
_{\mid \mathbf{T=0}}\mathbf{=}\nabla $ is imposed on $\tilde{R}$ we get
modifications which are equivalent to that for $f(R).$ In general, the
cycling/ ekpyrotic models with nontrivial $F$ are derived for certain
nonholonomically induced torsion configuration.

\subsection{Models with effective anisotropic fluids}

To understand what is the difference between a usual $f(R)$ theory and a
Finsler type one with $\tilde{R}(R)$ we have to consider physical equations
with N--elongated partial derivatives and certain information of osculating
approximation (\ref{osc}). For simplicity, we shall study a model of
effective locally anisotropic perfect fluid elaborated as follows. This
allows us to write directly generalizations of FLRW equations in N--adapted
form,
\begin{equation}
3H^{2}=\kappa ^{2}\tilde{\rho},\ 3H^{2}+2(\mathbf{e}_{t}H)=-\kappa ^{2}%
\tilde{p},  \label{flrweq}
\end{equation}%
where the energy--density and pressure of the locally anisotropic perfect
fluid is defined in such a way that for $\tilde{R}=\tilde{R}(R)$
\begin{eqnarray}
\tilde{\rho} &=&\kappa ^{-2}\{(\partial _{R}\tilde{R})^{-1}\left[ \frac{1}{2}%
\tilde{R}+3H\mathbf{e}_{t}(\partial _{R}\tilde{R})\right] -3(\mathbf{e}%
_{t}H)\},  \label{efd} \\
\tilde{p} &=&-\kappa ^{-2}\{(\partial _{R}\tilde{R})^{-1}\left[ \frac{1}{2}%
\tilde{R}+2H\mathbf{e}_{t}(\partial _{R}\tilde{R})+\mathbf{e}_{t}(\mathbf{e}%
_{t}(\partial _{R}\tilde{R}))\right] +(\mathbf{e}_{t}H)\}.  \notag
\end{eqnarray}%
We use such a N--adapted system of reference when (\ref{dder}) and (\ref%
{ddif}) are different in a form when pressure $\tilde{p}$ of such a "dark"
fluid is the same in all space--like directions (we remember that "tilde" is
used for certain values encoding contributions from a nontrivial $F$). The
equations of state (EoS) and EoS parameter in this Finsler model can be
written/computed respectively
\begin{equation*}
\tilde{p}=-\tilde{\rho}+\ ^{1}\tilde{p}\mbox{ and }\tilde{w}=\tilde{p}/%
\tilde{\rho},
\end{equation*}%
where%
\begin{equation}
\ ^{1}\tilde{p}:=-\kappa ^{-2}\{4(\mathbf{e}_{t}H)+\mathbf{e}_{t}(\mathbf{e}%
_{t}\ln |\partial _{R}\tilde{R}|)+(\mathbf{e}_{t}\ln |\partial _{R}\tilde{R}%
|)^{2}]-H\mathbf{e}_{t}\ln |\partial _{R}\tilde{R}|\}  \label{eosc}
\end{equation}%
has to be defined from a combination of FLRW equations,
\begin{equation}
2\mathbf{e}_{t}H=\kappa ^{2}\times \ ^{1}\tilde{p}(H,\mathbf{e}_{t}H,\mathbf{%
e}_{t}(\mathbf{e}_{t}H),...).  \label{eosceq}
\end{equation}%
It should be noted that in local coordinates such a system transform in a
very cumbersome combination of functional and partial derivatives on $t,$
with nontrivial N--coefficients, which is quite difficult to be solved in
explicit form for some prescribed values $N_{i}^{a}.$ We have to introduce
additional frame/coordinate transforms and assumptions on $H(t)$ in order to
construct exact solutions for certain locally anisotropic "cosmic" functions
(\ref{eosc}).

If the time variable is written as $t(\tilde{R}(R)),$ it is possible to
construct solutions of (\ref{eosceq}) following the approach developed in
\cite{saez,nojiri3}. For a new variable $\widetilde{r}(t,y(t))=\mathbf{e}%
_{t}\ln |\partial _{R}\tilde{R}(t)|,$ we transform the last equation into
\begin{equation*}
\mathbf{e}_{t}\widetilde{r}+\widetilde{r}^{2}-H\widetilde{r}=2\mathbf{e}%
_{t}H,\mbox{ for }\mathbf{e}_{t}\widetilde{r}=\partial _{t}\widetilde{r}%
-N_{t}^{a}(t,y(t))\partial _{a}\widetilde{r}.
\end{equation*}%
To reconstruct the Hubble parameter for such locally anisotropic
configurations, we can consider typical examples with power--law or
oscillating solutions. \ For instance, we show how we could generate
oscillating Finsler configurations. We assume a particular behaviour when $%
\widetilde{r}=r+\ ^{1}r,$ for $r(t)$ being the solution of
\begin{equation}
\mathbf{\partial }_{t}r+r^{2}-Hr=2\mathbf{\partial }_{t}H  \label{aux1}
\end{equation}%
and a small value $\ ^{1}r$ is to be found from (neglecting terms $(\
^{1}r)^{2}$ and $N\times (\ ^{1}r)$)
\begin{equation}
\mathbf{\partial }_{t}(\ ^{1}r)+(\ ^{1}r)(2r-Hr)=n  \label{aux1a}
\end{equation}%
where $n:=N_{t}^{a}\partial _{a}(r-2H).$ For $n=0$ and $\ ^{1}r=0,$ a
solution of (\ref{aux1}) was found in \cite{nojiri3}, when for $%
r(t)=r_{0}\cos \omega _{0}t$ it was reproduced
\begin{equation*}
f(R)=\int dR\ \exp [-r_{0}\omega _{0}^{-1}t(R)].
\end{equation*}%
In the case of Finsler modified gravity, the oscillating solutions of (\ref%
{aux1}) and (\ref{aux1a}) can be expressed in the form $\widetilde{r}%
(t)=r_{0}\cos \omega _{0}t+r_{1}\cos \omega _{1}t$. This which results in
reconstructions of type
\begin{equation}
\tilde{R}(R)=\int dR\exp \left\{ -r_{0}\omega _{0}^{-1}\sin [\omega
_{0}t(R)]-r_{1}\omega _{1}^{-1}\sin [\omega _{1}t(R)]\right\} .  \label{aux2}
\end{equation}%
The function $t(R)$ is an inversion of $R(t)=12H^{2}+6\mathbf{\partial }%
_{t}H.$\footnote{%
A similar formula was derived in \cite{stavr3} for a model of osculating
Finsler cosmology with weak anisotropy.} This corresponds to cyclic
evolution reproduced both in $f(R)$ and/or EFG gravity theories. The
complexity of such solutions does not allow to obtain explicit forms of $%
\tilde{R}(R)$ contributions. Nevertheless, we can distinguish from a "pure" $%
f(R)$ theory and that with a mixed with $F,$ when $f(R,F)\simeq \tilde{R}(R)$%
: in oscillations of type (\ref{aux2}), there is the second term induced by
off--diagonal terms summarized in $n$ as a source of (\ref{aux1a}).

\section{Tangent Lorentz Bundle Cosmology}

\label{stbc}The goal of this section is to show how using $\tilde{R}(R)$ and
related $f(R,F)$ cosmological scenarios on Lorentz manifolds $\mathbf{V}$ \
we can derive canonical models of Finsler gravity and cosmology on $T\mathbf{%
V.}$ Via Sasaki lifts of metrics (\ref{dm}) we shall construct theories on
total bundle spaces.

\subsection{Ekpyrotic and little rip Finsler cosmology}

Let us analyze scenarios of ekpyrotic / cyclic Universe \cite%
{steinh1,steinh2} in our case derived from Finsler like modifications of GR.
In this subsection, the constructions will be performed on $hT\mathbf{V}$
components of tangent Lorentz bundles. Cyclic solutions solve the problems
of standard cosmological model and provide a more complete theory. In
ekpyrotic models, a scalar field is necessary for reproducing the cycling
behavior and the $f(R)$ gravity admits such phases in time evolution.
Because the $\tilde{R}(R) $ locally anisotropic gravity can be modelled as a
$f(R)$ theory, it is clear that both types of theories contain cyclic
configurations.

It is possible to reconstruct a \ canonical Finsler like model with a
phantom phase and free of future singularity when it is generated a little
rip cosmology similar to \cite{fram1,fram2,brevik1}. We can understand how
such models may contain locally anisotropic modifications following
arguments: Using the first formula in (\ref{flrweq}), $3H^{2}=\kappa ^{2}%
\tilde{\rho}$, the effective density function $\tilde{\rho}$ (\ref{efd}) and
EoS parameter $\tilde{w}=-1-2(\mathbf{e}_{t}H)/3H^{2},$ we find terms of
type $\mathbf{e}_{t}(\partial _{R}\tilde{R})$ and $\mathbf{e}_{t}H$ with
contributions from the N--connection coefficients. Big rip singularities (by
definition) occur in a finite time $t_{s}$ when $a(t)$ and the
energy--density divergences. In $f(R)$ models, this can be analyzed in local
coordinate frames. Finsler modifications $\tilde{R}(R)$ result in
N--elongated partial derivatives $\mathbf{e}_{t}$ and nonholonomic frames of
reference.

It is not clear if we can formulate a general condition for $\tilde{\rho}$
and $\tilde{R}(R)$ do avoid divergences. We can consider an enough but not
sufficient condition that $\tilde{R}(R)>0$ but an additional investigation
of contributions with $N_{i}^{a}$ for little rip. Using the reconstruction
techniques from previous section, \ we shall prove that there are Finsler
models of general acceleration, which do not contain future singularities
and drive to a stronger growth in time than evolutions for big rip
singularities. We can use the generating function $\mathbf{\tilde{P}}(\chi
)=\exp [4\tilde{\beta}e^{\tilde{\alpha}\chi }],\tilde{\alpha}=const,\tilde{%
\beta}=const,$ resulting in cycling factors of type (\ref{cyclfact}). A
similar function can be considered for different constants when we test the
"cycling--accelerating" system for trivial N--coefficients, $\mathbf{P}(\chi
)=\exp [4\beta e^{\alpha \chi }],\alpha =const,\beta =const.$ For small
times $t\ll \tilde{\alpha}$, there are reproduced de Sitter solutions and a
Hubble parameter approximated to a constant. We get states with $\tilde{w}%
<-1,$ for a locally anisotropic phantom phase without big rip singularity.

Let us show that a dissolution of bound structures (little rip) induced by a
Finsler structure can occur. Using expansions of exponential functions into
power series and solutions (\ref{geneq}) for certain values of $\tilde{R}$
and/or $R$ as a generating function for solutions of (\ref{gensol2}), we
compute three reconstruction models (see similar details in the beginning of
Section V of \cite{nojiri3} and references therein). The solutions depend on
what type of scalar curvature we use, $R(t)=12H^{2}+6\mathbf{\partial }%
_{t}H, $ or $\tilde{R}(t)=12H^{2}+6\mathbf{e}_{t}H,$ and can be written in
the form, {\small
\begin{eqnarray}
f(R)& =&\alpha ^{2}\left( c_{1}+c_{2}\sqrt{4R/\alpha ^{2}+75}\right) e^{%
\sqrt{R/12\alpha ^{2}+25/16}}  \notag \\
&\sim &\kappa _{1}R+\kappa _{2}R^{2}/\alpha ^{2}+\kappa _{3}R^{3}/\alpha
^{4}+...,  \label{sol5} \\
\tilde{R}(R) &=&\tilde{\alpha}^{2}\left( c_{1}+c_{2}\sqrt{4R/\tilde{\alpha}%
^{2}+75}\right) e^{\sqrt{R/12\tilde{\alpha}^{2}+25/16}}  \notag \\
&\sim &\tilde{\kappa}_{1}R+\tilde{\kappa}_{2}R^{2}/\tilde{\alpha}^{2}+\tilde{%
\kappa}_{3}R^{3}/\tilde{\alpha}^{4}+...,  \notag \\
f(\tilde{R}) &=&\check{\alpha}^{2}\left( c_{1}+c_{2}\sqrt{4\tilde{R}/\check{%
\alpha}^{2}+75}\right) e^{\sqrt{\tilde{R}/12\check{\alpha}^{2}+25/16}}
\notag \\
&\sim &\check{\kappa}_{1}\tilde{R}+\check{\kappa}_{2}\tilde{R}^{2}/\check{%
\alpha}^{2}+\check{\kappa}_{3}\tilde{R}^{3}/\check{\alpha}^{4}+...,  \notag
\end{eqnarray}%
} where $c_{1}=-24\exp (-39/12)$ and $c_{2}=2\sqrt{3}$ are taken in order to
obtain the same approximations if $\tilde{R}\rightarrow R$ and $%
f(...)\rightarrow $GR; constants of type $\kappa _{i},\tilde{\kappa}_{i},...$
depend on $c_{1},c_{1}$ and respectively on $\alpha ,\tilde{\alpha}$ etc.

The above formulas with series decompositions distinguish three possible
cycling universes with rip evolution. In all cases, the quadratic terms on
curvature allow to cure the singularities. But approximations are different:
in the first case, we can obtain a recovering of GR form a "standard" $f(R)$
gravity theory; in the second case, we model Finsler modifications as a $%
f(...)$ theory without much information on Finsler generating function $F;$
in the third case, the rip evolution starts from a Finsler modified
spacetime. $\ $We suppose that such different cosmological scenarios can be
verified experimentally. For instance, \ for a Sun--Earth system with
densities of type $\rho =\rho _{0}e^{2\alpha t},\tilde{\rho}=\tilde{\rho}%
_{0}e^{2\tilde{\alpha}t},...,$ and $t_{0}=13.7$ Gyrs, we can obtain three
different approximations for the time of little rip (decoupling) $\ ^{LR}t,$%
{\small
\begin{equation*}
\ ^{LR}t =13.7\mbox{ Gyrs }+29.9/\alpha , \ ^{LR}\tilde{t} = 13.7%
\mbox{ Gyrs
}+29.9/\tilde{\alpha},\ \ ^{LR}\widehat{t} =13.7\mbox{ Gyrs }+29.9/\check{%
\alpha}.
\end{equation*}
} For oscillating solutions of type (\ref{aux2}), we get possible resonant
behaviour and shifting of decoupling etc. In general, models with $f(R),%
\tilde{R}(R),f(\tilde{R})$ possess different little rip properties.

\subsection{Reproducing a canonical model of EFG}

A metric compatible Finsler model on $T\mathbf{V}$ can be completely defined
by an action
\begin{equation*}
S=\int d^{4}x\delta ^{4}y\sqrt{|g_{ij}h_{ab}|}\left[ \frac{f(\ ^{F}R)}{2%
\widetilde{\kappa }^{2}}+\ ^{m}\widetilde{L}\right] .
\end{equation*}
This would result in functional dependencies of type $\ ^{F}R=\ ^{F}R(R)$
for a canonical scalar Finsler curvature (\ref{riccifs}). It is not clear
how we could extract information on a generalized gravitational constant(s)
and matter field interactions via $\ ^{m}\widetilde{L}$ in extra
"velocity/momentum" type dimensions. Nevertheless, we can encode such
contributions into certain polarized cosmological constants derived for
certain very general parameterizations of possible matter interactions with
"velocity/accelertion" variables and their duals. A "recovering" of Finsler
cosmology observational data on $\mathbf{V}$ to $T\mathbf{V}$ can be
performed following such a procedure for two distinguished cases:

\begin{enumerate}
\item Certain models of Finsler gravity can be associated to modified
dispersion relations\footnote{%
for $y^{i}=dx^{i}/d\tau ,$ when $x^{i}(\tau )$ is for a real parameter $\tau
$; $\rho _{\widehat{i}_{1}\widehat{i}_{2}...\widehat{i}_{2r}}(x)$ are
parameterized by 3--d space coordinates with "hats" on indices running
values $\widehat{i}=1,2,3$}
\begin{equation*}
\omega ^{2}=c^{2}[g_{\widehat{i}\widehat{j}}k^{\widehat{i}}k^{\widehat{j}%
}]^{2}(1-\frac{1}{r}\rho {_{\widehat{i}_{1}\widehat{i}_{2}...\widehat{i}%
_{2r}}y^{\widehat{i}_{1}}...y^{\widehat{i}_{2r}}}/{[g_{\widehat{i}\widehat{j}%
}y^{\widehat{i}}y^{\widehat{j}}]^{2r}}),  \label{disp}
\end{equation*}%
where a corresponding frequency $\omega $ and wave vector $k_{i}$ are
computed locally when the local wave vectors $k_{i}\rightarrow p_{i}\sim
y^{a}$ are related to variables $p_{i}$ which are dual to "fiber"
coordinates $y^{a}.$ These relations can be associated to a nonlinear
quadratic element (see details in \cite{vmdrhl,vcosm1,vcosm2}), {\small
\begin{equation}
ds^{2}=F^{2}(x,y)\approx -(cdt)^{2}+g_{\widehat{i}\widehat{j}}(x^{k})y^{%
\widehat{i}}y^{\widehat{j}}[1+\frac{1}{r}\frac{q_{\widehat{i}_{1}\widehat{i}%
_{2}...\widehat{i}_{2r}}(x^{k})y^{\widehat{i}_{1}}...y^{\widehat{i}_{2r}}}{%
\left( g_{\widehat{i}\widehat{j}}(x^{k})y^{\widehat{i}}y^{\widehat{j}%
}\right) ^{r}}]+O(\rho ^{2}).  \notag
\end{equation}
} For physical applications related to "small" deformations of GR, we can
consider that $g_{ij}=(-1,g_{\widehat{i}\widehat{j}}(x^{k}))$ in the limit $%
q \rightarrow 0$ correspond to a metric on a (pseudo) Riemannian manifold.

\item Finsler variables with a generating function ~$F(x,y)$ can be
introduced in GR and $f(R)$ modifications nonholonomic with 2+2 splitting.
Such a $F$ can be partially recovered in cosmological models using
observational data on $h$--subspace. To understand possible physical
implications of theories on $T\mathbf{V}$ is important to construct exact
solutions for locally anisotropic black holes, brane trapping/warping and
anisotropic cosmological solutions \cite{vfbr,vcosm3}. The generic
off--diagonal metrics, in both cases 1 and 2, are for Sasaki lifts (\ref{dm}%
).
\end{enumerate}

On tangent bundles, all fields depend on coordinates $u^{\alpha
}=(x^{i},y^{a})$ and the scalar curvature for the canonical d--connection $%
\mathbf{D}$ transform into $\tilde{R}\rightarrow $ $\ ^{F}R=\tilde{R}+\breve{%
R}$ (\ref{riccifs}). Having recovered the spacetime metric $h\mathbf{g=\{}$ $%
g_{ij}(t)=g_{ij}(t,y(t))\}$ for a cosmological model, we can construct up to
frame transform a metric (\ref{dm}),%
\begin{equation}
\mathbf{g}=h\mathbf{g}\oplus v\mathbf{g}=\ g_{ij}(t)\ e^{i}\otimes e^{j}+\
h_{ab}(x,y)\ \mathbf{e}^{a}\otimes \mathbf{e}^{b},  \label{dmc}
\end{equation}%
for certain coefficients $h_{ab}=e_{\ a}^{a^{\prime }}(x,y)e_{\
b}^{b^{\prime }}(x,y)g_{a^{\prime }b^{\prime }}(t)$ and $\mathbf{e}^{a}$
being determined by a canonical N--connection $\tilde{N}_{i}^{a}(x,y)$
induced via a chosen $F(x,y),$ see footnote \ref{fnotenc}. For cosmological
models, the coefficients of $v$--metric can be transformed via frame
transform to $h_{ab}(t,y).$ In general, such models are inhomogeneous on
fibre coordinates $y^{a}.$ Extending the scalar fields $\chi
(x^{i})\rightarrow \chi (x^{i},y^{a}),$ we can construct values of type (\ref%
{functa}), $\mathbf{\ }^{F}\mathbf{P[}\chi (\ ^{F}R)]>0$ and $\mathbf{\ }^{F}%
\mathbf{Q[}\chi (\ ^{F}R)],$ when $\ ^{F}R(R)=\ ^{F}\mathbf{P}[\chi (R)]R+\
^{F}\mathbf{Q[}\chi (R)].$

Actions of type (\ref{actf1}) can be generalized in the form%
\begin{equation*}
S=\int d^{4}x\delta ^{4}y\sqrt{|g_{ij}||h_{ab}|}\left[ \frac{1}{2\kappa ^{2}}%
\ ^{F}R(R,\chi )+\ ^{m}L\right] ,
\end{equation*}%
where $\delta y^{a}=\mathbf{e}^{a}$ (\ref{ddif}). We do not consider a
factor $e^{-2\chi }$ in this action because we shall work with another type
of conformal transforms when the action is re--defined in Finsler
generalized Einstein frames in order to remove strong coupling. On $T\mathbf{%
V,}$ a quintessence locally anisotropic action is postulated in the form%
\begin{equation*}
S=\int d^{4}x\delta ^{4}y\sqrt{|\widehat{g}_{ij}||\widehat{h}_{ab}|}\left[ \
^{F}R-\frac{1}{2}\omega (\chi )(\widehat{\mathbf{D}}_{\alpha }\chi )(%
\widehat{\mathbf{D}}^{\alpha }\chi )-U(\chi )\right] ,
\end{equation*}%
where 
$\widehat{\mathbf{g}}_{\alpha \beta }=\mathbf{\ }^{F}\mathbf{P}(\chi )%
\mathbf{g}_{\alpha \beta },\ \omega =12\left( \partial _{\chi }\sqrt{\mathbf{%
\ }^{F}\mathbf{P}}\right) ^{2}/\mathbf{\ }^{F}\mathbf{P,\ }U=\mathbf{\ }^{F}%
\mathbf{Q/(\mathbf{\ }^{F}\mathbf{P})}$ 
and $\widehat{\mathbf{D}}$ is the conformal transform of $\mathbf{D}.$ For
cosmological models, there are used (in our case, Finsler analogs) the
Jordan frames with%
\begin{equation*}
\widehat{a}(\widehat{t})=\sqrt{\mathbf{\ }^{F}\mathbf{P}(\chi (t))}a(t),%
\mbox{ \ for \ }d\widehat{t}=\sqrt{\mathbf{\ }^{F}\mathbf{P}(\chi (t))}dt,
\end{equation*}%
and locally anisotropic configurations $\chi (t,y^{a}),$ where the fiber
coordinates $y^{a}$ can be parameterized $y^{a}(t).$ In EFG, the evolutions
with $\widehat{a}(\widehat{t})$ and $\widehat{a}(t)$ are with respect to
N--adapted frames (\ref{dder}) and (\ref{ddif}).

We reproduce cycling universes with rip evolution of type (\ref{sol5})
prescribing\footnote{%
we extend on $T\mathbf{V}$ similar formulas (59) in \cite{nojiri3}; the
coefficients are redefined with "hats" and fixed in such a form on $h$%
--subspaces we get little rips as in $f(R);$ nevertheless, the parametric
dependence is modified on velocity like coordinates and parameters for EFG}
\begin{equation*}
\omega (\chi )=4\widehat{\alpha }\widehat{\beta }^{2}e^{2\widehat{\alpha }%
\chi }\mbox{ \ and \ }U(\chi )=-6\widehat{\alpha }^{2}(3+4\widehat{\beta }e^{%
\widehat{\alpha }\chi })(3+8\widehat{\beta }e^{\widehat{\alpha }\chi })\exp
(-4\widehat{\beta }e^{\widehat{\alpha }\chi }).
\end{equation*}%
Keeping only terms with $t$--evolution and $e^{\widehat{\alpha }t}\sim 1+%
\widehat{\alpha }t+O(t^{2})$ for small $t,$ the solution for scaling factors
are respectively constructed%
\begin{eqnarray}
\widehat{a}(t) &=&a_{0}e[6(\widehat{\beta }e^{\widehat{\alpha }\chi }+%
\widehat{\alpha }t)],\mbox{ for }a_{0}=const\mbox{ and }\widehat{t}%
=\int_{-\infty }^{2\widehat{\beta }\exp [\widehat{\alpha }\chi ]}e^{z}/z;
\label{scalingf} \\
\widehat{a}(\widehat{t}) &=&\widehat{a}_{0}\widehat{t}^{6}\exp [6\widehat{%
\alpha }\widehat{\beta }e^{-2\widehat{\beta }}\widehat{t}],\mbox{ for }%
\widehat{a}_{0}=const\mbox{ and }\widehat{H}(\widehat{t})=\widehat{\alpha }%
\widehat{\beta }e^{-2\widehat{\beta }}+6/\widehat{t}.  \notag
\end{eqnarray}%
Exact solutions in EFG with such scaling factors will be constructed in next
section. Here we observe that the Universe with Einstein--Finsler frames
describe both a type of (initial) big bang singularity and a
super--accelerating evolution. This is a manifestation of scenarios with
little rip determined by possible locally anisotropic character of
gravitational interactions on $T\mathbf{V.}$ It seems that singularities can
be removed in Jordan frames adapted to N--connections. We conclude that
Finsler configurations extended on tangent Lorentz bundles may result in
dissolution of bound structures of certain classes of FLRW models originally
defined in GR and then extended to EFG.

\subsection{Cosmological solutions in Finsler gravity}

On $T\mathbf{V}$ of a Lorentz manifold $\mathbf{V,}$ the Einstein--Finsler
equations (\ref{cdeins}) for the canonical d--connection $\mathbf{D}$ are
for a 8--d spacetime endowed with nontrivial N--connection structure. Such
systems of nonlinear partial derivative equations (PDE) can be solved in
very general forms using the anholonomic frame method \cite{vfbr,vmdrhl,veym}%
. \ Locally anisotropic and Finsler like solutions in 4--d models of gravity
were constructed in Refs. \cite{vcosm2,vcosm3,vgontsa}. In this subsection,
we provide several examples of 8-d exact solutions which possess cyclic /
ekpyrotic and little rip properties.

\subsubsection{Decoupling of EFG cosmological equations}

We label local coordinates $x^{i}=(x^{1}=r,x^{2}=\theta
),y^{a_{1}}=(y^{3}=t,y^{4}=\varphi );$ $y^{a}=(y^{a_{2}},y^{a_{3}}),$ where
indices run respective values $a_{1}=3,4;a_{2}=5,6;a_{3}=7,8.$ We consider $%
(x^{i},y^{a_{1}})$ as coordinates on a 4-d Lorentz manifold $\mathbf{V,}$
and the coordinates $y^{a}$ as fiber coordinates in $T\mathbf{V.}$ \ This
reflects a conventional $2+2+2+2$ splitting of coordinates which will give
us the possibility to integrate the gravitational field equations "shell by
shell" increasing dimensions by 2. The FLRW cosmological solution can be
written in the form%
\begin{equation}
\ \ h\mathbf{g}\mathbf{=}a^{2}(t)\left( \frac{dr{\otimes }dr}{1-\kappa r^{2}}%
+r^{2}d\theta {\otimes }d\theta \right) -dt{\otimes }dt+a^{2}(t)r^{2}\sin
^{2}\theta d\varphi {\otimes }d\varphi ,  \label{frw}
\end{equation}%
with $\sigma =0,\pm 1.$ This metric is an exact solution of the Einstein
equations with a perfect fluid stress--energy tensor, $T_{\ \ \beta
}^{\alpha }=diag[-p,-p,\rho ,-p],$where $\rho $ and $p$ are the proper
energy density and pressure in the fluid rest frame. For simplicity, we
consider as a "prime" a $h$--metric ansatz containing conformal transform of
(\ref{frw}), multiplying on $a^{-2}$, with $\sigma =0,$ and in Cartesian
coordinates, $h\mathbf{\mathring{g}}=\mathring{g}_{i}dx^{i}{\otimes dx}^{i}+%
\mathring{h}_{a}dy^{a}{\otimes dy}^{a},$ where $\mathring{g}_{i}=1,\mathring{%
h}_{3}(y^{3})=-a^{-2}(t),\mathring{h}_{4}=1.$ To construct 8--d cosmological
Finsler like solutions, we shall use the ansatz{\small
\begin{eqnarray}
\mathbf{g} &=&\eta _{i}(x^{i})\mathring{g}_{i}dx^{i}{\otimes dx}^{i}+\eta
_{a_{1}}(x^{i},t)\mathring{h}_{a_{1}}\mathbf{e}^{a_{1}}{\otimes }\mathbf{e}%
^{a_{1}}+  \notag \\
&&h_{a_{2}}(x^{k},t,y^{5})\mathbf{e}^{a_{2}}{\otimes }\mathbf{e}%
^{a_{2}}+h_{a_{3}}(x^{k},t,y^{a_{2}},y^{7})\mathbf{e}^{a_{3}}{\otimes }%
\mathbf{e}^{a_{3}},  \notag \\
\mathbf{e}^{a_{1}} &=&dy^{a_{1}}+N_{i}^{a_{1}}(x^{k},t)dx^{i},
\label{ansatz1} \\
\mathbf{e}^{a_{2}}
&=&dy^{a_{2}}+N_{i}^{a_{2}}(x^{k},t,y^{5})dx^{i}+N_{a_{1}}^{a_{2}}(x^{k},t,y^{5})dy^{a_{1}},
\notag \\
\mathbf{e}^{a_{3}}
&=&dy^{a_{3}}+N_{i}^{a_{3}}(x^{k},t,y^{a_{2}},y^{7})dx^{i}+  \notag \\
&&N_{a_{1}}^{a_{3}}(x^{k},t,y^{a_{2}},y^{7})dy^{a_{1}}+N_{a_{2}}^{a_{3}}(x^{k},t,y^{a_{2}},y^{7})dy^{a_{2}},
\notag
\end{eqnarray}%
} where $h_{a_{1}}=\eta _{a_{1}}(t)\mathring{h}_{a_{1}}$ are defined by
polarization functions $\eta _{a_{1}}=1+\varepsilon \chi
_{a_{1}}(x^{i},y^{3}),$ $\varepsilon <1.$ This class of generic
off--diagonal metrics depend on time like coordinate, $t$, and on
fiber--like ones, $y^{a_{2}}$ and $y^{7},$ and possess a Killing symmetry on
$\partial /\partial y^{7}$ because the coefficients do not contain the
coordinate $y^{7}.$ In Cartesian coordinates, the $h$--part of ansatz (\ref%
{ansatz1}) is modified by $\eta _{i}=1+\varepsilon \chi _{i}(x^{i}),$ when $%
g_{i}=\eta _{i}\mathring{g}_{i}=e^{\psi (x^{i})}$.\footnote{%
We note that it is possible to construct more general classes of solutions
without Killing symmetries, nonhomogeneous cosmological metrics etc. Such an
ansatz is a natural one when on the 4--d base spacetimes the cosmological
metrics depend only on $t$ and can be modified by 3--d space coordinates
and/or several fiber type coordinates.}

We do not have explicit observational/experimental evidences to determine
what kind of sources $\mathbf{\Upsilon }_{\beta \delta }$ may have physical
importance for models of matter field interactions in the total space $T%
\mathbf{V} $. From a formal point of view, we can extend a
geometric/variational formalism for deriving energy--momentum tensors on $%
\mathbf{V}$ (for instance, for scalar, spinor, gauge etc fields) to
construct similar values using Sasaki lifts for metrics and adapting the
constructions to N--elongated frames (\ref{dder}) and (\ref{ddif}). For
simplicity, we shall approximate such possible "extra velocity"
contributions by matter interactions with an effective cosmological constant
$\Lambda $ when $\mathbf{\Upsilon }_{\beta }^{\alpha }=\Lambda \mathbf{%
\delta }_{\beta }^{\alpha }.$ Then we shall see what kind of 8-d generic
off--diagonal cosmological solutions would possess cyclic, ekpyrotic and/or
little rip properties in the h--part of the total metric. As a matter of
principle, such an effective cosmological constant can be "polarized" and
depends on base and fiber coordinates. Nevertheless, it can be transformed
into a constant value using re--definition of generating functions for
various classes of solutions as we shall prove below. There are also
possible certain mechanisms of Finsler brane trapping/warping of extra
"velocity" type coordinates when N--adapted geometric constructions are
considered \cite{vfbr,vmdrhl}.

The gravitational field equations (\ref{cdeins}) for the ansatz (\ref%
{ansatz1}) decouple in this form:\

In 4--d, we get a system of nonlinear PDEs which with respect to N--adapted
frames is written
\begin{eqnarray}
\partial _{11}^{2}\psi +\partial _{22}^{2}\psi &=&\Lambda ,  \label{eq1} \\
\phi ^{\ast }(\ln |h_{4}|)^{\ast } &=&\Lambda h_{3},  \label{eq2} \\
\beta N_{i}^{3}+\alpha _{i} &=&0,  \label{eq3} \\
(N_{i}^{4})^{\ast \ast }+\gamma (N_{i}^{4})^{\ast } &=&0,  \label{eq4}
\end{eqnarray}%
with $\phi ^{\ast }=\partial _{t}\phi ,\partial _{1}=\partial
_{x^{1}},\partial _{11}^{2}=\partial _{x^{1}x^{1}}^{2}$, where the
coefficients
\begin{equation}
\gamma =(\ln |h_{4}|^{3/2}-\ln |h_{3}|)^{\ast },\ \alpha _{i}=h_{4}^{\ast
}\partial _{i}\phi \mbox{ and }\beta =h_{4}^{\ast }\phi ^{\ast },
\label{coeff}
\end{equation}%
are determined by $h_{3}$ and $h_{4}$ via generating function
\begin{equation}
\phi (x^{i},t)=\ln |2(\ln \sqrt{|h_{4}|})^{\ast }|-\ln \sqrt{|h_{3}|}.
\label{genf}
\end{equation}

The equations on the first 2--d shell (with coordinates $y^{5}$ and $y^{6}$)
are
\begin{eqnarray}
(\partial _{5}\ ^{1}\phi )\partial _{5}(\ln |h_{6}|) &=&\Lambda h_{5},
\label{eq2a} \\
\ ^{1}\beta N_{i}^{5}+\ ^{1}\alpha _{i} &=&0,\ ^{1}\beta N_{a_{1}}^{5}+\
^{1}\alpha _{a_{1}}=0,  \label{eq3a} \\
\partial _{55}^{2}(N_{i}^{6})+\ ^{1}\gamma \partial _{5}(N_{i}^{6})
&=&0,\partial _{55}^{2}(N_{a_{1}}^{6})+\ ^{1}\gamma \partial
_{5}(N_{a_{1}}^{6})=0,  \label{eq4a}
\end{eqnarray}%
for coefficients
\begin{eqnarray}
\ ^{1}\gamma &=&\partial _{5}(\ln |h_{6}|^{3/2}-\ln |h_{5}|),\ \ ^{1}\alpha
_{i}=(\partial _{5}h_{6})(\partial _{i}\ ^{1}\phi ),  \label{coef1} \\
\ ^{1}\alpha _{a_{1}} &=&(\partial _{5}h_{6})(\partial _{a_{1}}{}^{1}\phi
),\ ^{1}\beta =(\partial _{5}h_{6})(\partial _{5}\ ^{1}\phi ),  \notag
\end{eqnarray}%
determined by $h_{5}$ and $h_{6}$ via
\begin{equation}
\ ^{1}\phi (x^{k},t,y^{5})=\ln |2\partial _{5}(\ln \sqrt{|h_{6}|})|-\ln
\sqrt{|h_{5}|}.  \label{genf1}
\end{equation}%
In similar form, using coefficients $h_{7}$ and $h_{8}$ (and third "shell
coordinates" $y^{7}$ and $y^{6}$) we obtain {\small
\begin{eqnarray}
(\partial _{7}\ ^{2}\phi )\partial _{7}(\ln |h_{8}|) &=&\Lambda h_{7},
\label{e2b} \\
\ ^{2}\beta N_{i}^{7}+\ ^{2}\alpha _{i} &=&0,\ ^{2}\beta N_{a_{1}}^{7}+\
^{2}\alpha _{a_{1}}=0,\ ^{2}\beta N_{a_{2}}^{7}+\ ^{2}\alpha _{a_{2}}=0,
\label{e3b} \\
\partial _{77}^{2}(N_{i}^{8})+\ ^{2}\gamma \partial _{7}(N_{i}^{8})
&=&0,\partial _{77}^{2}(N_{a_{1}}^{8})+\ ^{2}\gamma \partial
_{7}(N_{a_{1}}^{8})=0,  \label{e4b} \\
\partial _{77}^{2}(N_{a_{2}}^{8})+\ ^{2}\gamma \partial _{7}(N_{a_{2}}^{8})
&=&0,  \notag
\end{eqnarray}%
} for coefficients
\begin{eqnarray}
\ ^{2}\gamma &=&\partial _{7}(\ln |h_{8}|^{3/2}-\ln |h_{7}|),\ \ ^{2}\alpha
_{i}=(\partial _{7}h_{8})(\partial _{i}\ ^{2}\phi ),  \label{coef2} \\
\ ^{2}\alpha _{a_{1}} &=&(\partial _{7}h_{8})(\partial _{a_{1}}{}^{2}\phi
),\ ^{2}\alpha _{a_{2}}=(\partial _{7}h_{8})(\partial _{a_{2}}{}^{2}\phi ),\
^{2}\beta =(\partial _{7}h_{8})(\partial _{7}\ ^{2}\phi ),  \notag
\end{eqnarray}%
generated by
\begin{equation}
\ ^{2}\phi (x^{k},t,y^{a_{2}},y^{7})=\ln |2\partial _{7}(\ln \sqrt{|h_{8}|}%
)|-\ln \sqrt{|h_{7}|}.  \label{genf2}
\end{equation}

We can see that the system of equations (\ref{eq2})-- (\ref{eq4}) is
similar, respectively, to (\ref{eq2a})-- (\ref{eq4a}) and (\ref{e2b})-- (\ref%
{e4b}). Such equations can be integrated consequently by adding additional
dependencies on next shell coordinates.

\subsubsection{Generating off--diagonal cosmological Finsler solutions}

The $h$--metric is given by $e^{\psi (x^{k})}dx^{i}\otimes dx^{j}$, where $%
\psi (x^{k})$ is a solution of (\ref{eq1}) considered as a 2--d Laplace
equation (\ref{eq1}). It depends on effective cosmological constant $\Lambda
.$

We can integrate the system (\ref{eq2}) and (\ref{genf}), for $\phi ^{\ast
}\neq 0.$ Such a condition can be satisfied by choosing a corresponding
system of frames/coordinates; it is possible to construct solutions choosing
$\phi $ with $\phi ^{\ast }=0,$ as particular cases (for simplicity, we omit
such considerations in this paper). Defining $A:=(\ln |h_{4}|)^{\ast }$ and $%
B=\sqrt{|h_{3}|},$ we re--write
\begin{equation}
\phi ^{\ast }A=\Lambda B^{2},\ Be^{\phi }=2A.  \label{aux6e}
\end{equation}%
If $B\neq 0,$ we get $B=(e^{\phi })^{\ast }/2\Lambda $ as a solution of a
system of quadratic algebraic equations. This formula can be integrated on $%
dt$ which results in
\begin{equation}
h_{3}=\ ^{0}h_{3}(1+(e^{\phi })^{\ast }/2\Lambda \sqrt{|\ ^{0}h_{3}|})^{2}.
\label{solh3}
\end{equation}%
Introducing this $h_{3}$ in (\ref{aux6e}) and integrating on $t,$ we get the
coefficients
\begin{equation}
h_{4}=\ ^{0}h_{4}\exp [\frac{e^{2\ \phi }}{8\Lambda }],  \label{solh4}
\end{equation}%
for an integration function $\ ^{0}h_{4}=$ $\ ^{0}h_{4}(x^{k}).$ We can fix $%
\ ^{0}h_{a}=$ $\mathring{h}_{a}$ as in (\ref{ansatz1}).

Having defined $h_{a}$ we can compute the N--connection coefficients as
solutions of (\ref{eq3}) and (integrating two times on $t)$ (\ref{eq4}),%
\begin{eqnarray}
w_{i} &=&-\partial _{i}\phi /\phi ^{\ast },  \label{ncoeff} \\
n_{k} &=&\ ^{1}n_{k}+\ ^{2}n_{k}\int dt\ h_{3}/(\sqrt{|h_{4}|})^{3},  \notag
\end{eqnarray}%
for integration functions $\ ^{1}n_{k}(x^{i})$ and $\ ^{2}n_{k}(x^{i}).$

Introducing solutions (\ref{solh3}), (\ref{solh4}) and (\ref{ncoeff}) for
the $h$--metric of ansatz (\ref{ansatz1}), we get an quadratic element for
nonhomogeneous 4-d cosmologies, {\small
\begin{eqnarray}
&&ds^{2}=e^{\psi }(dx^{i})^{2}+\mathring{h}_{3}(1+\frac{(e^{\phi })^{\ast }}{%
2\Lambda \sqrt{|\ \mathring{h}_{3}|}})^{2}\ [dt-\frac{\partial _{i}\phi }{%
\phi ^{\ast }}dx^{i}]^{2}+  \label{gensol1} \\
&&\mathring{h}_{4}\exp [\frac{e^{2\ \phi }}{8\Lambda }][dy^{4}+(\
^{1}n_{k}+\ ^{2}n_{k}\int dt\frac{h_{3}}{(\sqrt{|h_{4}|})^{3}})dx^{i}]^{2}.
\notag
\end{eqnarray}%
} The solutions depend on generating functions $\phi (x^{i},t)$ and $\psi
(x^{k})$ and on integration functions $\ \ ^{1}n_{k}(x^{k}),\
^{2}n_{k}(x^{k}).$ We approach the FLRW metric (\ref{frw}) if we chose such
values that $\eta _{i}=1+\varepsilon \chi _{i}(x^{i})\rightarrow 1$ and $%
\eta _{a_{1}}=1+\varepsilon \chi _{a_{1}}(x^{i},y^{3})\rightarrow 1$ and the
N--connection coefficients vanish\footnote{%
Here we note that the FLRW solution is for perfect fluid stress--energy
tensor and not for a cosmological constant. Nevertheless, using generating
and integration functions for coupled nonlinear systems we can define limits
to necessary type diagonal metrics via corresponding nonholonomic
constraints and frame transforms.}. This class of modified Finsler like
spacetimes (effectively modelled for 2+2 decompositions) are characterized
by a nontrivial torsion field determined only the coefficients of metric
(and respective N--connection). So, extra--dimensional Finsler contributions
can be via off--diagonal extension of solutions, nonlinear polarizations of
physical constants and metric coefficients and induced torsion. Such metrics
(being generic off--diagonal) can not be diagonalized via coordinate
transforms because the anholonomy coefficients in (\ref{anhrel}) are not
zero for arbitrary generating/ functions.

Metric of type (\ref{gensol1}) can be constrained additionally in order to
construct exact solutions for the Levi--Civita connection $\nabla .$ We have
to consider solutions with $\ ^{2}n_{k}=0,\partial _{i}(\
^{1}n_{k})=\partial _{k}(\ ^{1}n_{i})$ and $w_{i}=-\partial _{i}\phi /\phi
^{\ast }$ and $h_{4}$ subjected to
\begin{equation*}
w_{i}^{\ast }=\mathbf{e}_{i}\ln |h_{4}|,\partial _{i}w_{j}=\partial
_{j}w_{i},n_{i}^{\ast }=0,
\end{equation*}%
see details how to solve such equations in \cite{vexsol1,vfbr,vmdrhl,veym}
(this is possible, for instance, for separation of variables). Even for
solutions with $\nabla ,$ we get only effective Einstein spaces with locally
anisotropic polarizations and off--diagonal terms induced from Finsler
gravity.

Let us consider extra shell fiber contributions. The equations (\ref{eq2a}%
)-- (\ref{eq4a}) and (\ref{e2b})-- (\ref{e4b}) can be integrated following
above procedure but for corresponding coefficients, (\ref{coef1}) and (\ref%
{coef2}), and generating functions, (\ref{genf1}) and (\ref{genf2}). Such a
class of 8--d solutions are parameterized by the quadratic element {\small
\begin{eqnarray}
&&ds^{2}=e^{\psi }(dx^{i})^{2}+\mathring{h}_{3}(1+\frac{\partial
_{t}(e^{\phi })}{2\Lambda \sqrt{|\ \mathring{h}_{3}|}})^{2}\ [dt-\frac{%
\partial _{i}\phi }{\phi ^{\ast }}dx^{i}]^{2}  \label{gensol3} \\
&&+\mathring{h}_{4}\exp [\frac{e^{2\ \phi }}{8\Lambda }][dy^{4}+(\
^{1}N_{k}^{4}+\ ^{2}N_{k}^{4}\int dt\frac{h_{3}}{(\sqrt{|h_{4}|})^{3}}%
)dx^{i}]^{2}  \notag \\
&&+\ ^{0}h_{5}(1+\frac{\partial _{5}e^{\ ^{1}\phi }}{2\Lambda \sqrt{|\ \
^{0}h_{5}|}})^{2}\ [dy^{5}-\frac{\partial _{i}(\ ^{1}\phi )}{\partial _{5}(\
^{1}\phi )}dx^{i}-\frac{\partial _{a_{1}}(\ ^{1}\phi )}{\partial _{5}(\
^{1}\phi )}dy^{a_{1}}]^{2}  \notag \\
&&+\ ^{0}h_{6}\exp [\frac{e^{2\ \ ^{1}\phi }}{8\Lambda }][dy^{6}+(\
^{1}N_{k}^{6}+\ ^{2}N_{k}^{6}\int dy^{5}\frac{h_{5}}{(\sqrt{|h_{6}|})^{3}}%
)dx^{i}  \notag \\
&&+(\ ^{1}N_{a_{1}}^{6}+\ ^{2}N_{a_{1}}^{6}\int dy^{5}\frac{h_{5}}{(\sqrt{%
|h_{6}|})^{3}})dy^{a_{1}}]^{2}+\ ^{0}h_{7}(1+\frac{\partial _{7}e^{\
^{2}\phi }}{2\Lambda \sqrt{|\ \ ^{0}h_{7}|}})^{2}\   \notag \\
&&[dy^{7}-\frac{\partial _{i}(\ ^{2}\phi )}{\partial _{7}(\ ^{2}\phi )}%
dx^{i}-\frac{\partial _{a_{1}}(\ ^{2}\phi )}{\partial _{7}(\ ^{2}\phi )}%
dy^{a_{1}}-\frac{\partial _{a_{2}}(\ ^{2}\phi )}{\partial _{7}(\ ^{2}\phi )}%
dy^{a_{2}}]^{2}  \notag \\
&&+\ ^{0}h_{8}\exp [\frac{e^{2\ \ ^{2}\phi }}{8\Lambda }][dy^{8}+(\
^{1}N_{k}^{8}+\ ^{2}N_{k}^{8}\int dy^{7}\frac{h_{7}}{(\sqrt{|h_{8}|})^{3}}%
)dx^{i}  \notag \\
&&+(\ ^{1}N_{a_{1}}^{8}+\ ^{2}N_{a_{1}}^{8}\int dy^{7}\frac{h_{7}}{(\sqrt{%
|h_{8}|})^{3}})dy^{a_{1}}+(\ ^{1}N_{a_{2}}^{8}+\ ^{2}N_{a_{2}}^{8}\int dy^{7}%
\frac{h_{7}}{(\sqrt{|h_{8}|})^{3}})dy^{a_{2}},  \notag
\end{eqnarray}%
}where there are used the integration functions $\ ^{1}N_{k}^{4},\
^{2}N_{k}^{4}$ depending on $(x^{i},y^{3}=t);$ $\ \ ^{0}h_{a_{1}},\
^{1}N_{k}^{6},\ ^{2}N_{k}^{6},\ ^{1}N_{a_{1}}^{6},\ ^{2}N_{a_{1}}^{6}$
depending on $(x^{i},y^{a_{1}},y^{5});$ $\ ^{0}h_{a_{2}},\ ^{1}N_{k}^{8},\
^{2}N_{k}^{8},\ ^{1}N_{a_{1}}^{8},\ ^{2}N_{a_{1}}^{8},\ ^{1}N_{a_{2}}^{8},\
^{2}N_{a_{2}}^{8}$ depending on $(x^{i},y^{a_{1}},y^{a_{2}},y^{7}).$ The
generating functions are
\begin{equation*}
\phi (x^{i},t),\partial _{t}\phi \neq 0;\ ^{1}\phi
(x^{i},y^{a_{1}},y^{5}),\partial _{5}\ ^{1}\phi \neq 0;\ ^{2}\phi
(x^{i},y^{a_{1}},y^{a_{2}},y^{7}),\partial _{7}\ ^{2}\phi \neq 0.
\end{equation*}%
We can fix such configurations which model certain scenarios in cosmology
and/or gravitational and matter field interactions.

\subsubsection{Extracting realistic cosmological configurations in EFG}

Metrics of type (\ref{gensol3}) define off--diagonal exact solutions for
8--d locally anistoropic generalizations of FLRW cosmology. In general, it
is not clear what kind of physical significance such nonhomogeneous
solutions may have. Playing with values and parameters in generating and
integration functions and source $\Lambda $, we can mimic different
scenarios and compare them, for instance, with modifications derived for
other types of gravity theories.

We note that we can generate off--diagonal solutions depending only on $t$
and fiber coordinates $(y^{a_{2}},y^{7})$ if via frame/coordinate transforms
and a corresponding fixing of generating/integration functions when the
coefficients of (\ref{gensol3}) do not depend on coordinates $(x^{i}).$ In
such cases, a series of coefficients $N_{i}^{a}=0$ and, the integration
functions are of type $\phi (t),\partial _{t}\phi \neq 0;\ ^{1}\phi
(t,y^{5}),\partial _{5}\ ^{1}\phi \neq 0;\ ^{2}\phi
(t,y^{5},y^{6},y^{7}),\partial _{7}\ ^{2}\phi \neq 0.$ The nonholonomic/
nonlinear gravitational dynamics on fiber variables may mimic scalar and
other matter field interactions in observable 4--d spacetime. It is possible
to apply various trapping/warping mechanisms like in brane gravity \cite%
{vfbr,vmdrhl} or using osculating approximations in Finsler gravity.

Introducing a cycling factor $\tilde{a}(t)=\exp [H_{0}t+b(t)]$ (\ref%
{cyclfact}) \ instead of $a(t)$ in \ $\mathring{h}_{3}(y^{3})=-a^{-2}(t),$
we can model such a cycling scenarios with respect to N--elongated frames (%
\ref{dder}) and (\ref{ddif}) when the nontrivial coefficients $N_{i}^{a}$
are functions on some variables from the set $(t,y^{5},y^{6},y^{7}).$ For
small values of $N$ (assuming that fiber gravity results only in small
corrections), we can put certain symmetries and boundary conditions on a
subclass of solutions (\ref{gensol3}), which possess cycling behavior and
may limit in 4-d the FLRW solution (\ref{frw}) but contains also certain
nontrivial anisotropic polarizations caused by possible Finsler like
interactions. For anisotropic spacetimes, we have to recover not a $f(R)$
theory and certain "exotic" states of matter but choosing a corresponding
set generating/integration functions.

To reproduce little rip evolution by off--diagonal Finsler metrics we can
introduce a scaling factor $\widehat{a}(t)=a_{0}e[6(\widehat{\beta }e^{%
\widehat{\alpha }\chi }+\widehat{\alpha }t)]$ (\ref{scalingf}) instead of $%
a(t)$ in \ $\mathring{h}_{3}(y^{3})=-a^{-2}(t),$ or try to get such a term
in $h_{3}$ and $h_{4},$ for (\ref{gensol3}). All geometric constructions
will evolve with respect to certain nonholonomic frames of reference, with
small corrections by N--coefficients.

Finally we note that a solution of type (\ref{gensol3}) is not written in
well known Finsler form (\ref{dm}) with standard Hessian $\tilde{g}_{ab}$ (%
\ref{hess}) and N--connection structure $\tilde{N}_{i}^{a}.$ Nevertheless,
we can related the coefficients of both representations using frame
transforms of type $g_{a^{\prime }b^{\prime }}=e_{\ a^{\prime }}^{a}e_{\
b^{\prime }}^{b}\tilde{g}_{ab}$ if a fundamental Finsler function $F$ is
taken from certain experimental data or fixed following certain theoretical
arguments. It is not convenient to find exact solutions working directly
with data $\left( \tilde{g}_{ab},\tilde{N}_{i}^{a}\right) $ because the
gravitational field equations contain in such cases forth order derivatives
on $F.$ To derive a general geometric method of constructing exact solutions
is possible for ansatz of type (\ref{gensol3}) with general dependence on
certain classes of generating/integration functions. Then constraining the
integral varieties of solutions and after corresponding frame transforms and
distorting of connections (\ref{dcdc}) we can derive exact solutions for
necessary types of connections, symmetries, cosmological evolution behavior
which may \ explain cyclic and little rip properties etc.

\section{Discussion and Conclusions}

\label{sconcl}

The procedure for explicit reconstruction of modified gravity using
cosmological observation data \cite{nojiri3,nojiri4,saez} was generalized
for a comparative study of $f(R)$ and Einstein--Finsler gravity (EFG)
theories. We have shown how such models can be extended around general
relativity (GR) action and various classes of cosmological solutions. The
conclusion on existence of cyclic evolution scenarios driven by Finsler
fundamental functions was made using locally anisotropic variants of
scalar--tensor gravity theories. Nonholonomic constraints and off--diagonal
metrics may mimic matter field interactions and evolution processes.
Geometrical actions, off--diagonal nonlinear dynamics and the conditions
usually considered for inflation models were shown that may lead to cyclic
universes and ekpyrotic effects.

We proved that both $f(R)$ and EFG theories encode scenarios with little rip
universe with dark energy represented by nonholonomically induced
non--singular phantom cosmology. So, such classed of theories are consistent
with observational/ experimental data and present a realistic alternative to
the $\Lambda$CDM model. Various types of non--singular super--accelerating
and/or locally accelerated universes can be reconstructed in modified
gravity theories. The solutions presented here can be extended to Finsler
like brane and/or Ho\v rava--Lifshitz theories \cite{vfbr,vmdrhl}. Such
theories modelled as GR plus corrections pass various local tests.

It is a very complicate technical problem to construct generic off--diagonal
solutions in GR and modifications. We generalized our anholonomic frame
method \cite{vexsol1,veym} of constructing exact solutions in a form which
would allow to decouple and solve in off--diagonal forms the gravitational
field equations for modified theories. Following such an approach, it is
possible to generate anisotropic off--diagonal cosmological scenarios with
dependence on "velocity-momentum" type coordinates $y^{a},$ see metrics (\ref%
{gensol3}). Such models can be considered as low energy limits of Finsler
stochastic metrics of type (3.4), (3.10) and (4.18) in \cite{mavr2}, when  $%
y^{a}$ can be associated to momentum transfers by quantum-gravitational
fluctuations in the space-time and D--brane/--particle foam. It is possible
to provide a microscopic background following this approach.  Our class of
solutions allow us to model cosmological evolution scenarios with
acceleration and dark energly/matter effects in a form adapted to the
nonlinear connection structure (with is very important in Finsler gravity).
For various types of Finsler gravity and cosmology theories with $y^{a}$
treated as extra dimension coordinates, the physical meaning of such
variables is model dependent. For instance, we can not perform
compactification of velocity type coordinates because there is a constant
speed of light and nontrivial N--connection structure. It is important to
find  solutions with trapping/warping effects.  Having constructed certain
physically important  Finsler gravity/cosmology models we can perform the
osculating approximation $y(x)$ to metrics of type  (\ref{osc}). This allow
us to compare observational data for different modified theories of gravity
and GR which in all case are considered in real/effective 4--d spacetime.\footnote{There were elaborated various geometric and physical models of Finsler spacetimes and applications  Finsler methods in modern gravity and cosmology, see \cite{stavr1,stavr2,mavr1,mavr2,sindoni,visser,vcosm3,vexsol1,veym,vfbr,bao,caponio} and references therein. A series of works are with metric noncompatible Finsler type connections, or without explicit assumptions and physical motivations on  Finsler nonlinear and linear connection structures. The main physical problems of such works with nonmetricity fields are  that we are not able to define analogs of Finsler type spinors,  Finsler--Dirac equations and  well--defined "standard" conservation laws. There  are not "simple" analytic methods for  constructing physically important exact solutions for Finsler type black holes, brane configurations and to  formulate a recovering procedure from cosmological data became very problematic  etc, see critical remarks in Refs. \cite{vcritic,vcosm2,vcosm1,vrflg,vgontsa}. Perhaps, most closed to general relativity and standard physics are the (generalized/modified) Finsler like models constructed on  tangent bundles to Lorentz manifolds when there are defined natural lifts of geometric/physical objects from Einstein spaces to  Einstein--Finsler analogs. In such cases,  there are canonical  metric compatible Finsler connections adapted to the nonlinear connection structure and an osculating approximation can be performed. Following such an approach,  it was formulated  a self-consistent axiomatic formalism \cite{vcosm2} which is very similar to  that for the general relativity theory. It is possible  to elaborate certain renormalizable models of Finsler like quantum gravity \cite{vmdrhl} and recovering procedures from cosmological data, to find exact solutions \cite{vfbr,vexsol1,veym} etc.}

Finally, we not that in this paper the reconstruction procedure is involved
in a new form when the cosmological models are determined by certain classes
of generating/integration functions. The cyclic/ekpyrotic/ little rip
universe scenarios are possible in EFG but it remains to understand how
natural and realistic are solutions with locally anisotropic effects on
tangent bundles on Lorentz manifolds. This is one of the scopes of our
future works.

\vskip5pt \textbf{Acknowledgements: } SV research is partially supported by
the Program IDEI, PN-II-ID-PCE-2011-3-0256. He is grateful to N. Mavromatos,  A. Kouretsis,
E. Elizalde and S. D. Odintsov for hospitality and discussions. PS is
supported by a Special Account (for Research Grants) of University of
Athens. He thanks M. Francaviglia for discussions.

\setcounter{equation}{0} \renewcommand{\theequation}
{C.\arabic{equation}} \setcounter{subsection}{0}
\renewcommand{\thesubsection}
{B.\arabic{subsection}}

\end{document}